\documentstyle[12pt,dina4p,epsfig]{article}
\title{
  \vskip-2cm
  {\baselineskip16pt
    \centerline{\normalsize \tt DESY 96-077 \hfill ISSN 0418-9833}
    \centerline{\normalsize \tt hep-ph/9605210 \hfill}
    \centerline{\normalsize \tt May 1996 \hfill}
  }
  \vskip2cm
  {\bf 
    Large Transverse Momentum Jet Production and DIS Distributions 
    of the Proton
  }
  \author{
    {M.\ Klasen,  G.\ Kramer} \\
    {II. Institut f\"ur Theoretische Physik}\thanks
    {Supported by Bundesministerium f\"ur Forschung und
     Technologie, Bonn, Germany under Contract 05\,6HH93P(5) and
     EEC Program "Human Capital and Mobility" through Network
     "Physics at High Energy Colliders" under Contract
     CHRX-CT93-0357 (DG12 COMA)} \\
     {Universit\"at Hamburg} \\
     {D - 22761 Hamburg, Germany}
    }
  \date{}
}
\begin{document}
\maketitle
\begin{abstract}
We have calculated the single jet inclusive cross section as measured
at Fermilab in next-to-leading order QCD using recent parton distributions
of the CTEQ collaboration. We studied the scheme dependence of the jet
cross section by employing the $\overline{\mbox{MS}}$ and DIS factorization
schemes consistently. For $E_T > 200$ GeV, we find that the cross section
in the DIS scheme is larger than in the $\overline{\mbox{MS}}$ scheme
yielding a satisfactory description of the CDF data over the whole $E_T$
range in the DIS scheme.
\end{abstract}
\vskip1cm
%
Recently, the CDF collaboration presented a precise measurement of the
inclusive differential cross section for jet production in $p\overline{p}$
collisions at 1.8 TeV \cite{xxx1}. The measurement was compared to
next-to-leading order (NLO) perturbative QCD predictions \cite{xxx2}
for jet transverse energies $E_T$ in the range of 15 to 440 GeV in the
central pseudorapidity region $0.1 \leq |\eta | \leq 0.7$, using a
selection of commonly used parton distributions. The experimental results
for $E_T > 200$ GeV show evidence of a possible deviation as compared to
the NLO prediction based on the current sets of parton distributions,
which are obtained from global analyses \cite{xxx3,xxx4,xxx5,xxx6} of deep 
inelastic
lepton-nucleon scattering and related data. Before explanations for this
deviation based on new physics \cite{xxx7} are seriously considered, it
is crucial to study possible explanations within the Standard Model. \\

Clearly, the perturbative predictions of the jet cross section depend on
the parton distributions. Then the question arises whether the parton
distributions can be adjusted to accommodate the jet measurements while
still having a good description of the data sets as used in previous
global analyses. This vital question has been addressed by two groups.
Huston et al. \cite{xxx8} have carried out a new global QCD analysis
incorporating the CDF inclusive jet data with $E_T > 75$ GeV (the very
precise data in the low $E_T$ range are excluded due to potential
theoretical and experimental problems) and reached the conclusion, that
there is enough flexibility in NLO global analyses to enhance the high
$E_T$ inclusive jet cross section by $25-35\% $ above previous
calculations. A similar analysis was done by Glover et al. \cite{xxx9}.
They also incorporated the CDF single jet inclusive measurement in a
global NLO parton analysis of the available deep inelastic and related
data. In particular, they included the jet data in the region $E_T > 50$
GeV in their fit. They find that it is impossible to accommodate both
the jet data over the complete $E_T$ range and the deep inelastic structure
function data. However, the CDF data for $E_T < 200$ GeV and the deep
inelastic data were reasonably well fitted simultaneously. A fit over the
full $E_T$ range yields quarks which are completely incompatible with the
large-$x$ structure function data, a value for $\alpha_s$, which is larger
than the current world average, and a renormalization of the CDF jet
data, which is barely compatible with the allowed range \cite{xxx1}. 
So from these two studies it still seems to be unclear, whether the
difference between the CDF inclusive jet cross section data and the NLO
predictions can be attributed to unsufficient knowledge of the parton
distributions. \\

It is well known that there are other uncertainties of the perturbative
calculations which should be considered. First, we have the dependence
on the renormalization and factorization scales. Second, changes in the
strong coupling $\alpha_s$ resulting in changes of $\Lambda_{\mbox{QCD}}$
must be considered. These uncertainties have been discussed in \cite{xxx8}
and mainly affect the renormalization and not the shape of the inclusive
jet cross section. Third, there exists the dependence of the perturbative
predictions on the factorization scheme. To our knowledge, this
dependence of the single jet cross section has not been investigated yet. \\

Beyond leading order, the parton distribution functions $f^a(x,\mu^2)$
depend on the factorization scheme. In applications to physical processes,
the scheme chosen for the parton distributions must match that for the
hard scattering cross section in the parton model formula \cite{xxx10}. The
same parton distributions in two different schemes differ because of two
effects. First, the scheme influences the evolution kernel. Second,
the functional dependence at the starting scale $\mu=Q_0$ differs since
the parton distributions are fitted to the same data with the evolution
kernel for the appropriate scheme. Thus, they are functionally equivalent
in the sense that they yield the same physical cross sections for the
data included in the analysis. The $\overline{\mbox{MS}}$ parton
distributions are guaranteed to satisfy the momentum sum rule. The DIS
scheme \cite{xxx11} was defined specifically to make the relation between
the parton distributions and $F_2^\gamma$  as simple as possible by
absorbing all the NLO terms into the definition of $f^q_{DIS}(x,\mu^2)$.
This does not yet define the DIS prescription for the gluon distribution.
Here, the convention has beed adopted to require the momentum sum rule
to be preserved in the DIS scheme as well. However, this requirement fixes
only the second moment of the gluon distribution, so that one completes
the definition by requiring the condition on the second moment to be valid
for all moments. This convention was first introduced by Diemoz et al.
\cite{xxx12} and is now generally applied for parton distribution function
analyses in the DIS scheme. \\

Functionally equivalent parton distributions in
the $\overline{\mbox{MS}}$ and the DIS schemes have been constructed by the
CTEQ collaboration \cite{xxx4,xxx5}. The most recent ones are the sets
CTEQ3M ($\overline{\mbox{MS}}$) and CTEQ3D (DIS), respectively \cite{xxx5}.
These two sets are obtained from independent fits to the same data sets
under the same assumptions except for calculating the evolution kernel. The
total $\chi^2$ for the two sets is supposed to be very similar.
Unfortunately, this is not stated explicitly in \cite{xxx5}. However,
this fact was reported for the earlier sets CTEQ1M and CTEQ1D \cite{xxx4}.
Thus, the two sets describe totally equivalent physics as defined by the
data sets selected in the analysis. Now we can ask ourselves whether they
also yield the same single jet inclusive cross sections or not. \\

There also exist DIS parton distributions from the other collaborations
\cite{xxx3,xxx6}. In particular, MRS(D0') and MRS(D-') sets in the DIS
scheme are available \cite{xxx3}. They are obtained from the corresponding
sets in $\overline{\mbox{MS}}$, which were fitted to the deep inelastic
and other data, by applying the O($\alpha_s$) perturbative transformation
formula between the two schemes. It is known that this may be unreliable
in situations where the NLO terms involving, for example, a large gluon
contribution, are of comparable size as the LO term involving small
sea quarks \cite{xxx5}. Furthermore, the DIS version of MRS is not
equivalent to the Diemoz et al. description, since only the quark
distribution functions are transformed. The GRV collaboration \cite{xxx6}
also constructed a DIS set in their recent 1994 analysis (GRV(94)) by
fitting $\overline{\mbox{MS}}$ and DIS parton distributions independently
to the deep inelastic and other data. Unfortunately, this analysis is
restricted to three flavors with the charm contribution to the deep
inelastic structure function generated perturbatively. This makes this
set less useful for calculations of other hard scattering cross sections.
Therefore, we shall restrict ourselves to the CTEQ3 sets in the following.
\\

Our calculation uses the NLO parton level program JETSAM \cite{xxx13} which
originally was designed for the calculation of inclusive single jet
cross sections in resolved photoproduction. This program was developed on
the basis of the theory of Aversa et al. \cite{xxx15} which uses the
phase space slicing method to cancel collinear divergences. The cuts and
the algorithm for defining jets out of up to three partons are modelled
as closely as possible to the experimental set-up. The jets are defined
according to the Snowmass algorithm \cite{xxx14} with a jet cone size of
$R = 0.7$ and lie in the pseudorapidity range between $0.1 \leq |\eta |
\leq 0.7$ as in the CDF analysis. The factorization and renormalization
scales are chosen to be $\mu_F=\mu_R=E_T/2$. The Monte Carlo JETSAM 
contains the necessary modifications of the NLO hard scattering cross
sections to run it in the $\overline{\mbox{MS}}$ and the DIS scheme as 
defined above. \\

First, we compare the CDF jet data with the NLO prediction obtained from
the CTEQ3M set of partons. The fractional difference between the data and
the theoretical prediction is plotted in Fig.~1. The data are taken from
Table 2 in \cite{xxx1}. Over the observed range of $E_T$, the experimental
cross section decreases by more than a factor $10^8$ and the quadratic
sum of the {\it correlated} systematic uncertainties grows from 
$\pm 18 \% $ at $E_T = 50$ GeV to $\pm 36 \% $ at $E_T = 400$ GeV. In Fig.~1,
the agreement of the data with the theory for the CTEQ3M set is measured
by the distance of the data points from the horizontal line at zero.
The normalization shown is absolute. These results show excellent agreement
in shape and normalization for $E_T < 200$ GeV, where the cross section
falls by six orders of magnitude. Above $E_T = 200$ GeV, the CDF cross
section is significantly higher than the NLO cross section. This agrees
with the statements made in ref. \cite{xxx1,xxx8,xxx9}, where other parton
sets in the $\overline{\mbox{MS}}$ scheme obtained from recent global
analyses had been selected for the NLO calculation of the jet cross section.
Thus, the agreement below 200 GeV and disagreement above 200 GeV seems
to be a unique feature of all $\overline{\mbox{MS}}$ parton sets from
recent global analyses which include both fixed-target and HERA deep
inelastic data. \newpage

\begin{figure}[hhh]
 \begin{center}
  \begin{picture}(15,9)
   \epsfig{file=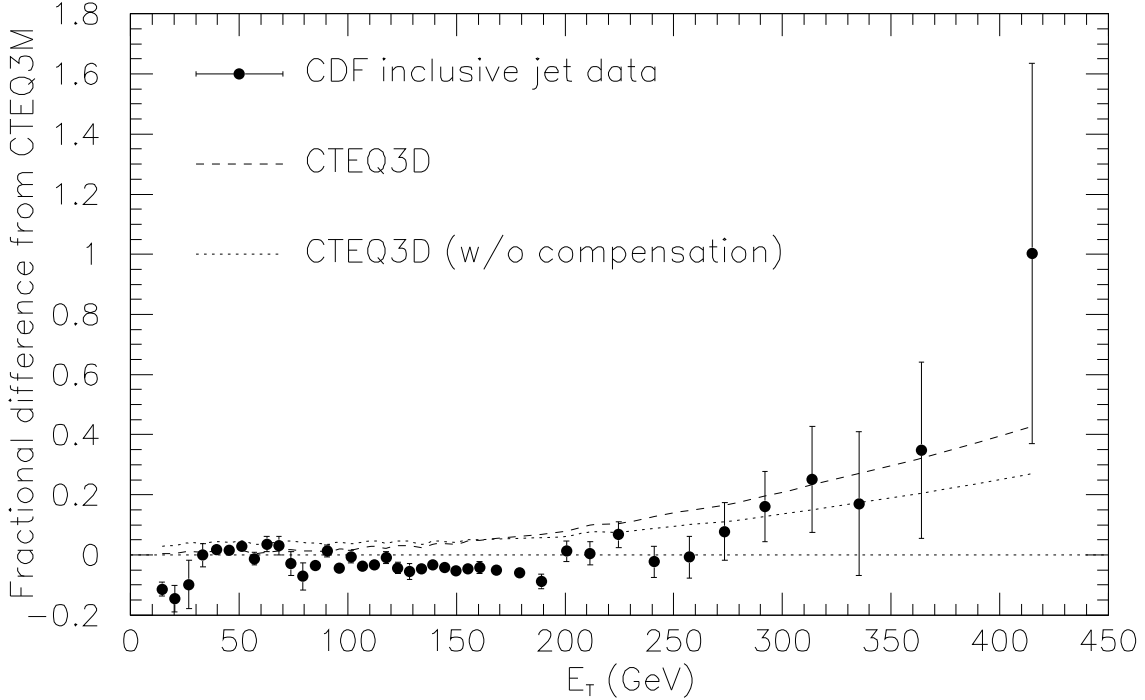,bbllx=66pt,bblly=274pt,bburx=493pt,bbury=540pt,%
           width=15cm,clip=}
  \end{picture}
  \caption{Fractional differences of the CDF inclusive jet data, the CTEQ3D 
           prediction, and the CTEQ3D prediction without compensation terms
           from the CTEQ3M NLO QCD predictions.}
 \end{center}
\end{figure}

In Fig.~1, we have also plotted the prediction for the 
CTEQ3D parton distributions. It is given by the dashed line which presents
the fractional difference of the CTEQ3D cross section from the CTEQ3M
cross section, i.e. (CTEQ3D-CTEQ3M)/CTEQ3M. As we can see, this line
deviates very little from the horizontal line (the CTEQ3M result) for
$E_T < 150$ GeV, showing that the single inclusive jet cross section does
not depend on the chosen factorization scheme in this $E_T$ range.
Above $E_T = 150$ GeV, the DIS prediction starts to deviate from
the horizontal line and increases up to 0.4 at $E_T = 400$ GeV. Thus,
in the large $E_T$ range the jet cross section depends on the factorization
scheme. For $E_T > 150$ GeV, the DIS cross section is larger than the
$\overline{\mbox{MS}}$ cross section so that it agrees with all the large
$E_T$ data points except the last one which, however, has a large
statistical error. In addition, we must take into account the systematic
error of the experimental data \cite{xxx1}. \\

In Fig.~1, we also show
a second curve (dotted). This presents the fractional difference to the
CTEQ3M prediction, obtained also with the DIS parton distribution set
CTEQ3D, but without the appropriate modification of the hard scattering
cross section to the DIS scheme, i.e. this cross section is left as
calculated in the $\overline{\mbox{MS}}$ scheme. Of course, this is
an inconsistent procedure. But it shows that the deviation of the DIS
cross section from the $\overline{\mbox{MS}}$ cross section at 
$E_T = 400$ GeV, which is $40\% $, comes to a large extent from the
modified parton distributions ($25\% $) and to a lesser extent from
the modified subtraction terms in the hard scattering cross section
($15\% $). At the larger values of $E_T$, these two effects do not compensate
each other as one might expect. At the smaller $E_T < 150$ GeV, this
compensation occurs, the dashed curve is nearer to the horizontal line
than the dotted one, although the difference between the two curves
is very small. \\

We also performed the same calculations for the older
sets CTEQ2M and CTEQ2D \cite{xxx5}, respectively. The results are similar.
The agreement between the data and the predictions with either the
$\overline{\mbox{MS}}$ or DIS parton distributions in the $E_T < 150$ GeV
is less satisfactory. The fractional difference of the CTEQ2D cross
section from the CTEQ2M cross section shows the same increase with
increasing $E_T$ up to the value 0.5 at $E_T = 400$ GeV.\\

Fig.~2 compares the quark and gluon distributions at $Q^2 = 10$ and
$10^4~\mbox{GeV}^2$ of the two sets CTEQ3M and CTEQ3D.
\begin{figure}[hhh]
 \begin{center}
  \begin{picture}(11,16)
   \epsfig{file=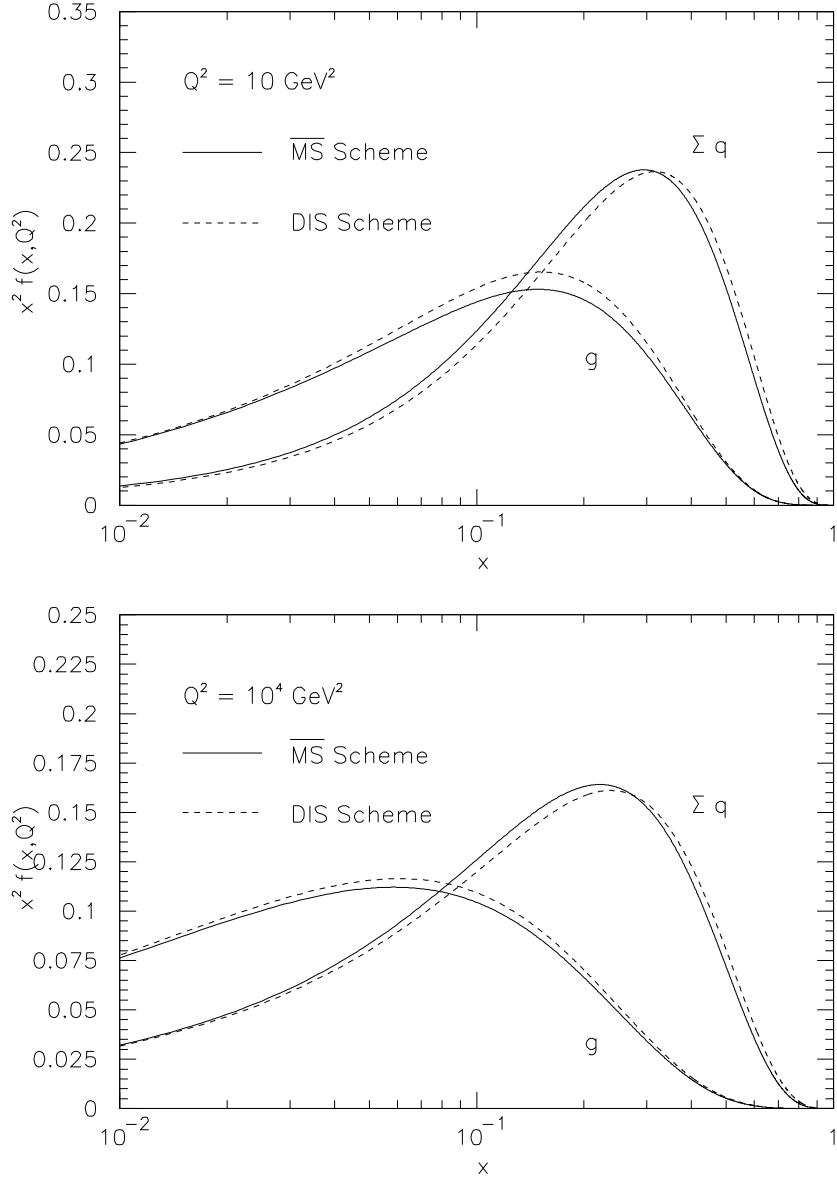,bbllx=55pt,bblly=100pt,bburx=484pt,bbury=710pt,%
           width=11cm,clip=}
  \end{picture}
  \caption{The CTEQ3M and CTEQ3D parton distributions
           $\sum_a (q_a+\overline{q}_a)$ and $g$ at $Q^2 = 10$ and 
           $10^4~\mbox{GeV}^2$.}
 \end{center}
\end{figure}
We plot
the singlet quark distribution $\sum_a (q_a +\overline{q}_a)$ 
and the gluon distribution $g$ multiplied
by $x^2$, so that the area under the curves is the total momentum
fraction carried by the quarks and gluons. We see that for both $Q^2$ 
values, the DIS quarks are larger (smaller) than the $\overline{\mbox{MS}}$
quarks for $x \geq 0.3 (x \leq 0.3)$, whereas the DIS gluons are larger
than the $\overline{\mbox{MS}}$ gluons over the whole $x$ range of
$0.01 < x < 1$. At smaller $E_T$, where the jet cross section receives
contributions mainly from $qg-$ and $gg-$initiated subprocesses \cite{xxx9},
the compensation between the change of the parton distributions from
$\overline{\mbox{MS}}$ to DIS and the modification of the hard scattering
subprocesses seems to work reasonably well (see Fig.~1). For $E_T > 200$
GeV, the jet production is dominated by $q\overline{q}$ initiated 
subprocesses \cite{xxx9}. Since centrally produced jets of transverse
energy $E_T$ sample partons at $x\simeq x_T = \frac{2E_T}{\sqrt{s}}$,
we can understand that the increased quark distribution for $x > 0.3$ in the
DIS scheme leads to an increased jet cross section for $E_T > 250$ GeV
in case that in this region the compensation is distorted due to the
singling out of a particular subprocess. The increase of the gluon
distribution in this region is small, so that the $q\overline{q}$ process
is mainly responsible for the increase of the jet cross section. For smaller
$E_T$, the effect of the larger gluon is compensated by a smaller quark 
distribution for $x < 0.3$. \\

In summary, we have studied the dependence of the inclusive jet cross
section on the factorization scheme. We have considered two schemes,
$\overline{\mbox{MS}}$ and DIS, for which parton distributions have
been constructed in a global analysis by the CTEQ collaboration. We find
that the inclusive jet cross section for $E_T > 200$ GeV in the DIS
scheme is up to $40\% $ larger than in the $\overline{\mbox{MS}}$ scheme
and accounts well for the CDF inclusive jet cross section data in this
$E_T$ region as well as in the low $E_T$ range.\\

Of course, the significant scheme dependence of the high $E_T$ jet
cross section is an artifact of finite order perturbation theory. In
infinite order, this dependence is absent. From this point of view, our
results indicate that the NLO theory is inadequate and one must go to the
NNLO in order at least to diminish the scheme dependence and to obtain
more reliable results. However, since parton distribution functions are
involved, which are determined from other processes, the problem of the
scheme dependence might have another solution. It is conceivable, that
the parton distributions in the DIS scheme are the right ones with
enough flexibility built in to describe all deep inelastic data together
with the CDF jet data. Then, the task would be to find the corresponding
$\overline{\mbox{MS}}$ set which describes the same data by perhaps
changing the input parametrization of the distribution functions, in
particular that of the gluon, which is much less constrained by deep
inelastic data than the quark distribution. Such an investigation is
beyond the intention of this work. 
We therefore conclude that it might still be possible that the
difference between the CDF inclusive data and the NLO QCD predictions
can be due to a deficiency in the way parton distributions in the
$\overline{\mbox{MS}}$ scheme are constructed in the usual global analyses
work as performed by the CTEQ, MRS, and GRV collaborations.

\end{document}